# Testing of Bridging Faults in AND-EXOR based Reversible Logic Circuits

Avik Chakraborty
avik@bhelhyd.co.in

*Abstract*—**Reversible circuits find applications in many areas of Computer Science including Quantum Computation. This paper examines the testability of an important subclass of reversible logic circuits that are composed of k-wire controlled NOT (k-CNOT with k ≥ 1) gates. A reversible k-CNOT gate can be implemented using an irreversible k-input AND gate and an EXOR gate. A reversible k-CNOT circuit where each k-CNOT gate is realized using irreversible k-input AND and EXOR gate, has been considered. One of the most commonly used Single Bridging Fault model (both wired-AND and wired-OR) has been assumed to be type of fault for such circuits. It has been shown that an (n+p)-input AND-EXOR based reversible logic circuit with p observable outputs, can be tested for single bridging faults (SBF) using $(3n + \lceil \log_2 p \rceil + 2)$ tests.**

*Keywords-Reversible, k-CNOT, Irreversible gate, Bridging Fault, Test Set*

## I. INTRODUCTION

Reversible circuits are classical counterparts of Quantum Circuits which are inherently reversible [1], [2]. It has manifold applications in low power CMOS quantum computing, nanotechnology, optical computing, computer graphics, DNA technology and cryptography. In quantum circuits, the unit of quantum information is a qubit, which can be either in a zero state ($|0>$) or a one state ($|1>$). It can also be in a state which is a superposition of these states, i.e. $\alpha|0> + \beta|1>$, where $\alpha$ and $\beta$ are complex numbers called amplitudes so that $|\alpha|^2 + |\beta|^2 = 1$. Quantum computation can solve exponentially hard problems (ex. Prime Factorization) in polynomial time by exploiting the superposition.

Various physical realizations of quantum gates for ex. trapped ion technology, Nuclear Magnetic Resonance (NMR), photons and non-linear optical media, cavity-quantum electro-dynamic devices, spin of electrons in semiconductor, have been reported. This paper investigates testability of reversible k-CNOT circuits ($k \geq 1$) where each k-CNOT gate has been realized using an irreversible k-input AND and an EXOR gate. As these gates are classical irreversible gates, a reversible k-CNOT circuit can be realized in silicon (for ex. CMOS). The fault model assumed is single Bridging Fault model where two adjacent lines can be physically shorted giving rise to wired-AND and wired-OR fault. However, as actual implementation of reversible circuits is unknown, the feasibility of the bridging fault model is yet to be established.

Though testability of reversible circuits has been reported in literature, the possible implementation of such circuits has not been considered. Testing of reversible logic circuits for Single and Multiple stuck-at faults has been investigated [3]. Universal Testability of k-CNOT circuits for stuck-at faults has also been addressed [4]. Polian et al [5] have proposed four new fault models namely *single missing gate, multiple missing gate, partial missing gate and repeated missing gate* fault models and investigated testability of reversible logic circuits under these faults. Zhong et al [6] have proposed a new fault model, *crosspoint fault model* for reversible circuits and given a heuristic method for test generation. Rahaman et al [7] have investigated the problem of bridging fault testing in reversible circuits. Universal Testability for input bridging faults has also been reported [8].

This paper is organized as follows. The assumed Single Bridging Fault (SBF) model has been elaborated in Section II. The AND-EXOR based realization of reversible k-CNOT circuit has been outlined in Section III. Section IV illustrates the test set generation procedure. Section V concludes the paper.

## II. FAULT MODEL

Bridging Faults are caused by physical shorts between two (or more) signal lines. All the lines involved in a bridging fault have the same logic value. Only Single Bridging Fault (SBF) between any pair of lines has been considered and it has been assumed that only a single pair of lines is faulty at any time. The logical effect of such a fault may be of classical AND or OR type, as described in Fig.1 and Table 1. Thus, a single bridging fault between two lines in a reversible circuit can be tested by an input vector *t* if and only if the affected lines are driven by opposite logic values and that error is propagated to an observable output of the circuit.

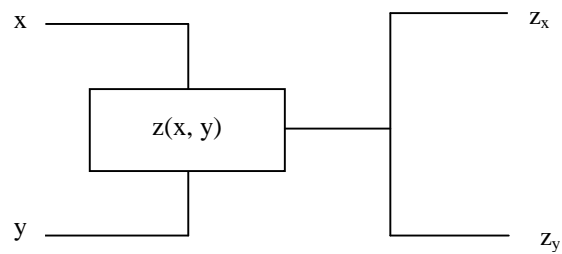

**Fig. 1**: Bridging Fault between two lines x and y

TABLE I: Logical Effect of Bridging Faults

| Input Values | | AND Bridging | | OR Bridging | |
|---|---|---|---|---|---|
| x | y | $z_x$ | $z_y$ | $z_x$ | $z_y$ |
| 0 | 0 | 0 | 0 | 0 | 0 |
| 0 | 1 | 0 | 0 | 1 | 1 |
| 1 | 0 | 0 | 0 | 1 | 1 |
| 1 | 1 | 1 | 1 | 1 | 1 |

## III. AND-EXOR BASED REALIZATION

A reversible k-CNOT gate can be realized by an irreversible k-input AND gate and an EXOR gate as shown in Fig.2. In this paper, we consider only those reversible circuits which are composed of k-CNOT gates where $k \geq 1$. A 0-CNOT gate is a simple NOT gate and it can be realized from a 1-CNOT gate where the control input is set to logical 1. By synthesizing the 0-CNOT gates from 1-CNOT gates any k-CNOT circuit can be converted to a circuit where $k \geq 1$.

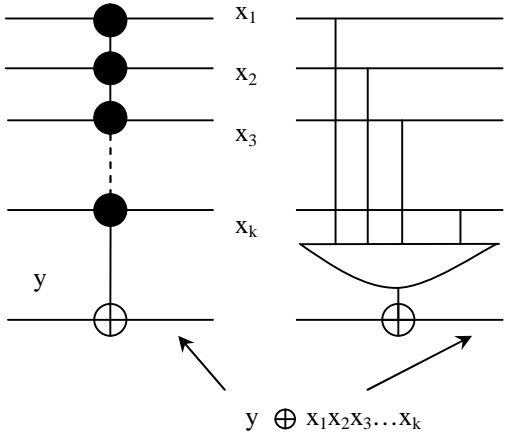

**Fig. 2**: Realization of a reversible k-CNOT gate

A reversible k-CNOT circuit is composed by first defining the input/output wires from the reversible function specification, and then concatenating reversible gates to some subset of the wires of the circuit. This gives the notion of levels in the circuit; the inputs are at level 0 and the outputs of any gate are at one plus the highest level of any of its inputs. The maximum level is $d$ which is equal to the number of gates in the circuit. Consider the reversible circuit shown in Fig. 3. It has two types of inputs, i.e. $x_i$s ($1 \leq i \leq n$) which are transmitted unchanged and $c_j$s ($1 \leq j \leq p$) which are acted upon by the EXOR gates to produce the outputs $f_j$s ($1 \leq j \leq p$) which are assumed to be the only observable outputs of the circuit. In this example, $n = 7$ and $p = 3$. It can be observed that each $f_j$ realizes a positive polarity Reed-Muller (PPRM) expression with $n$ inputs and it consists of only positive product terms (as $k \geq 1$).

$$f_j(x_1, x_2, ...., x_n) = \bigoplus_{i=0}^{2^n-1} a_i \mu_i \quad (1)$$

where $a_i \in \{0, 1\}$ and

$\mu_i = x_n^{e_n} x_{n-1}^{e_{n-1}} ..... x_2^{e_2} x_1^{e_1} = \Pi_{k=1}^{n} x_k^{e_k}$ where $e_k \in \{0, 1\}$ such that $e_n e_{n-1}...e_2 e_1$ is a binary number which equals $i$. Moreover, $x_i^0 = 1$ and $x_i^1 = x_i$. $\bigoplus$ denotes summation over GF(2), the Galois Field of two elements.

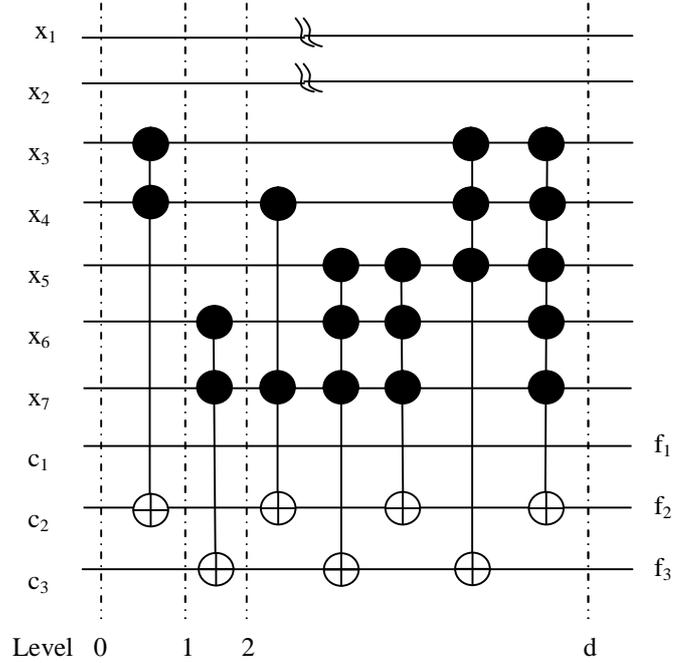

**Fig. 3**: A Reversible k-CNOT circuit with $n=7$ and $p=3$

The functions realized by the circuit shown in Fig.3 are as follows ($f_1$ is not shown in Fig. 3).

$$f_1(x_1, x_2, x_3, x_4, x_5, x_6) = x_1 \oplus x_2 \oplus x_4 \oplus x_5 \oplus x_1 x_2 \oplus x_1 x_2 x_3 \oplus x_1 x_5 \oplus x_2 x_6 \oplus x_3 x_4 \oplus x_3 x_5 \oplus x_1 x_2 x_4 \oplus x_1 x_2 x_3 x_4 x_5 \quad (2)$$

$$f_2(x_3, x_4, x_5, x_6, x_7) = x_3 x_4 \oplus x_4 x_7 \oplus x_5 x_6 x_7 \oplus x_3 x_4 x_5 x_6 x_7 \quad (3)$$

$$f_3(x_3, x_4, x_5, x_6, x_7) = x_6 x_7 \oplus x_5 x_6 x_7 \oplus x_3 x_4 x_5 \quad (4)$$

The AND-EXOR based realization of a generic reversible k-CNOT circuit is shown in Fig.4.

## IV. TEST SET GENERATION

Any two lines $h_i$ and $h_j$ in the circuit may be shorted resulting in either AND (($h_i h_j$)∗) or OR (($h_i h_j$)₊) bridging fault. Let $N_k(x_i)$ ($N_k(x_i x_j)$) denote the number of AND gates to which $x_i$ ($x_i x_j$) is an input in the output $f_k$, where, $1 \leq k \leq p$. $N_k(x_i \cup x_j)$ denotes number of AND gates to which $x_i$ or $x_j$ are inputs in $f_k$.

$$N_k(x_i \cup x_j) = N_k(x_i) + N_k(x_j) - N_k(x_i x_j) \quad (5)$$

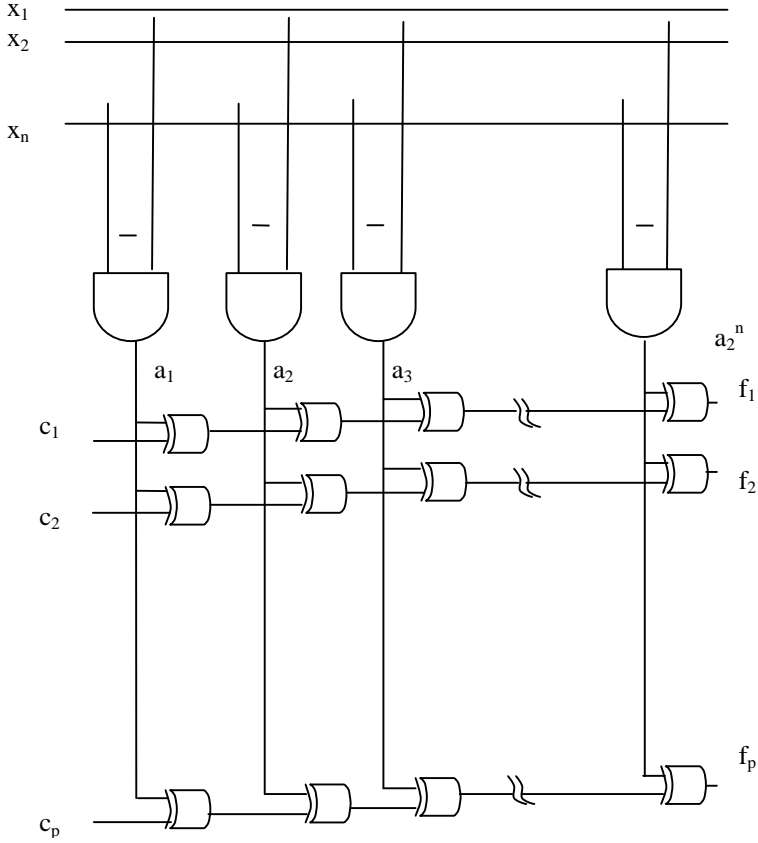

**Fig. 4**: AND-EXOR Realization of a k-CNOT circuit

The AND-EXOR based circuit has $(n+p)$ controllable inputs ($x_i$s for $1 \leq i \leq n$ and $c_j$s for $1 \leq j \leq p$) and $p$ observable outputs ($f_k$s for $1 \leq k \leq p$). In this paper, $x_i$s and its fan-out lines and $a_i$s and its fan-out lines are not distinguished. Any fault which occurs on these lines has to be propagated to one of the observable outputs, i.e. $f_k$s. There are four types of bridging faults that may occur in the circuit:

1. Faults inside EXOR gates.
2. Bridging Fault between any $x_i$ and $x_j$, i.e. $(x_ix_j)_*$ and $(x_ix_j)_+$.
3. Intra-level Bridging Fault in the EXOR part.
4. Bridging Fault between any $a_i$ and $a_j$, i.e. $(a_ia_j)_*$ and $(a_ia_j)_+$.

### A. Detecting Faults inside EXOR gates

An EXOR gate can be realized in many ways. Irrespective of its implementation, any fault which occurs inside EXOR gate can be detected at observable output $f_i$s if each EXOR gate is tested exhaustively, i.e. applying all four patterns to each 2-input EXOR gate. All EXOR gates in an AND-EXOR based k-CNOT circuit with $(n+p)$ inputs can be tested for this fault with a test set $T_1$. The size of the test set $T_1$ is four irrespective of $n$ and $p$. The test set $T_1$ is shown below.

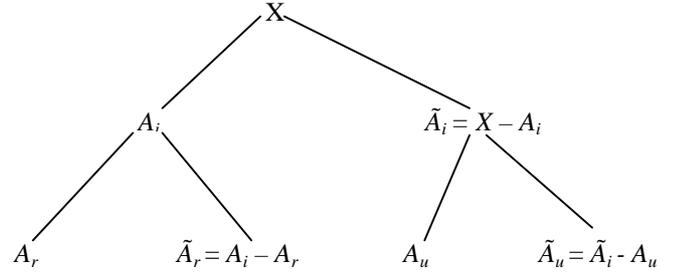

### B. Detecting $(x_ix_j)_*$

Let $A_i$ be the subset of input variables which are inputs to an AND gate whose output is $a_i$ (see Fig. 4) and cardinality of $A_i$ (number of variables in $A_i$) is $|A_i|$. Consider an AND gate with minimal $|A_i|$ and its input variables $x_i \in A_i$. Generate a test pattern $t_i$ such that $a_i(t_i) = 1$ and $x_j=0$ $\forall$ $x_j \notin A_i$. $t_i$ detects any BF $(x_ix_j)_*$ where $x_i \in A_i$ and $x_j \notin A_i$ since the output changes from '1' to '0' when the fault occurs. Test pattern $t_i$ divides the set of input variables $X = \{x_1x_2...x_n\}$ into two subsets $A_i$ and $\tilde{A}_i = X - A_i$ as shown in Fig. 5. Now the BFs to be detected are $(x_rx_s)_*$ and $(x_ux_v)_*$, $x_r$, $x_s \in A_i$ for all $r \neq s$ and $x_u$, $x_v \in \tilde{A}_i$ for all $u \neq v$.

```
                    X
                   / \
                  /   \
                 Ai    Ãi = X − Ai
                /\     /\
               /  \   /  \
              Ar  Ãr=Ai−Ar  Au  Ãu = Ãi - Au
```

**Fig. 5**: Binary Tree

The process of test generation is carried out by considering minimal subset of variables $A_r \subset A_i$ such that $(x_rx_s)_*$ are detected for all $x_r \in A_r$ and for all $x_s \in \tilde{A}_r = A_i - A_r$. This is accomplished by considering an AND gate with $A_r$ as its subset of inputs (and if required some other variables from $\tilde{A}_i$) such that the output of this AND gate $a_j(t_j) = 1$ for a test vector $t_j$ and $a_j(t_j) = 0$ if there is a bridging fault $(x_rx_s)_*$ for all $x_r \in A_r$ and for all $x_s \in \tilde{A}_r$. Let's name the test set so constructed as $T_2$. The following example illustrates the construction of $T_2$ for AND-EXOR based reversible circuit shown in Fig.3.

Consider the expression for $f_1$, eqn. (2). Let us select four minimal $A_i$s, $x_1$, $x_2$, $x_4$ and $x_5$ from eqn. (2). The corresponding test cases are $t_1(1000000)$, $t_2(0100000)$, $t_3(0001000)$ and $t_4(0000100)$. Then consider the product term $x_3x_4$ from eqn. (3). The test case is $t_5(0011000)$. Similarly, the product term is $x_4x_7$ and test case is $t_6(0001001)$. These 6 test cases detect all $(x_ix_j)_*$ faults. The corresponding binary tree is shown in Fig. 6. The test set $T_2$ so constructed is shown below. As $c_i$s are don't cares, they are not shown.

$$T_2 \begin{array}{c} \\ t_1= \\ t_2= \\ t_3= \\ t_4= \\ t_5= \\ t_6= \end{array} \begin{array}{ccccccc} x_1 & x_2 & x_3 & x_4 & x_5 & x_6 & x_7 \\ \left[\begin{array}{ccccccc} 1 & 0 & 0 & 0 & 0 & 0 & 0 \\ 0 & 1 & 0 & 0 & 0 & 0 & 0 \\ 0 & 0 & 0 & 1 & 0 & 0 & 0 \\ 0 & 0 & 0 & 0 & 1 & 0 & 0 \\ 0 & 0 & 1 & 1 & 0 & 0 & 0 \\ 0 & 0 & 0 & 1 & 0 & 0 & 1 \end{array}\right] \end{array}$$

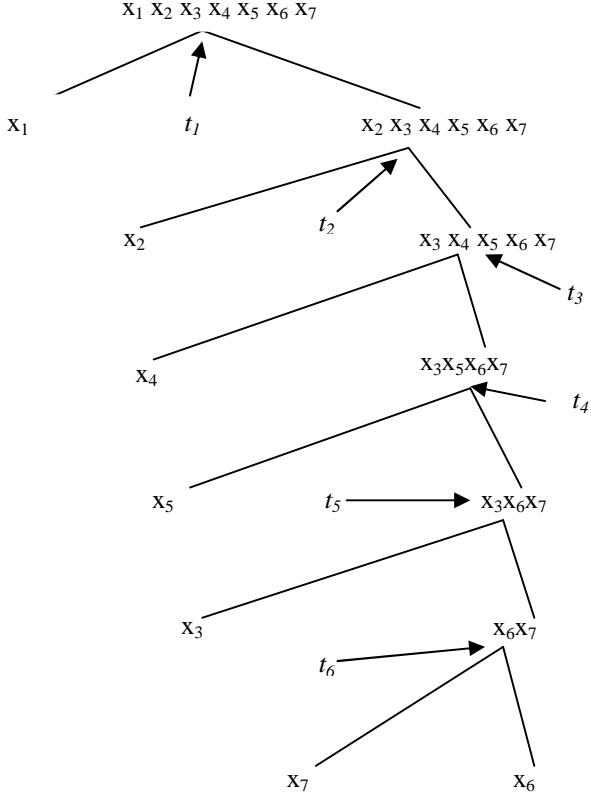

**Fig. 6**: Binary Tree for detection of $(x_i x_j)_*$

C. *Detecting $(x_i x_j)_+$*

To generate a test set for detection of $(x_i x_j)_+$, an [nXn] parity matrix $P = \{p_{ij}\}$ will be constructed where the $i^{th}$ row and the $j^{th}$ column correspond to the variables $x_i$ and $x_j$, respectively and $p_{ij}$ is given by

$$p_{ij} = \begin{cases} 1 & \text{if any of } N_k(x_i x_j) \text{ is odd for } 1 \leq k \leq p \\ 0, & \text{otherwise} \end{cases} \quad (6)$$

where $N_k(x_i x_i) = N_k(x_i)$, (see eqn.(5)). Please note that the $P$ matrix gives information on whether a variable $x_i$ is going to an even or an odd number of AND gates depending on $p_{ij}$ being '0' or '1'. Similarly, it also gives information on every pair of variables $x_i$ and $x_j$, whether they jointly go to an even or an odd number of AND gates. To detect a bridging fault, the fault should affect an odd number of gates so that its effect will result in a change of any $f_k$. The generation of test patterns for is described below depending on various cases whether $p_{ii} = 0$ or 1 and $p_{ij} = 0$ or 1.

Case (a): $p_{ii} = 1$. Then, the test pattern

$t = (x_1, x_2, \dots x_{i-1}, x_i, x_{i+1}, \dots x_n)$

$= (1\ 1\ \dots\ 1\ 0^i\ 1\ \dots\ 1\ 1)$ (7)

detects $(x_i x_j)_+$, for all $j \neq i$. Input set X is partitioned into $\{x_i\}$ and $X-\{x_i\}$, and $t \in T_5$ where $T_5$ is constructed later. Test patterns for all variables $x_i \in X$ for which $p_{ii} = 1$ should be generated first before going to case (b) or case (c).

Case (b): $p_{ii} = 0$. If $x_k$ is such that $p_{ik} = 1$ then test pattern

$t = (x_1, x_2, \dots x_i, \dots x_k, \dots x_n)$

$= (1\ 1\ \dots\ 1\ 0^i\ 1\ \dots\ 1\ 0^k\ 1\ \dots 1\ 1)$ (8)

detects $(x_i x_j)_+$ for all $j \neq i, k$ since the number of AND gates enabled after bridging is equal to $N_k(x_i) - N_k(x_i x_k) =$ odd. If $P_{kk} = 0$, then the same test pattern detects $(x_k x_j)_+$ for all $j \neq i, k$. If $P_{kk} = 1$, then it comes under case (a).

If case (a) and case(b) are not satisfied for a single variable or a number of variables, the test patterns for them will be generated as described in case (c).

Case (c): $P = [0]$. Find a sub function $f_i = f(x_1, x_2, \dots, x_{i-1}, 0, x_{i+1}, \dots, x_n)$ and the corresponding parity matrix $P_1$ for this function. If $P_1 = [0]$ for $f(X / x_i = 0)$ for all $x_i$, then consider $P_2$ for $f(X / x_i = x_j = 0)$; If $P_2 \neq [0]$ for some $x_i$ and $x_j$, then the test patterns are generated as in case (a) and (b). If $P_2 = [0]$, for all $x_i = x_j = 0$, consider $P_3$ for the case when three variables are set to 0 and so on.

Consider the circuit shown in Fig. 3 and equations (2), (3), (4). Let's name the test set so constructed as $T_3$. The $N_k(x_i x_j)$ values are shown below.

| $N_k(x_i x_j)$ | $f_1$ | $f_2$ | $f_3$ | $N_k(x_i x_j)$ | $f_1$ | $f_2$ | $f_3$ |
|---|---|---|---|---|---|---|---|
| $N(x_1 x_1)$ | 6 | 0 | 0 | $N(x_3 x_4)$ | 2 | 2 | 1 |
| $N(x_1 x_2)$ | 4 | 0 | 0 | $N(x_3 x_5)$ | 2 | 1 | 1 |
| $N(x_1 x_3)$ | 2 | 0 | 0 | $N(x_3 x_6)$ | 0 | 1 | 0 |
| $N(x_1 x_4)$ | 2 | 0 | 0 | $N(x_3 x_7)$ | 0 | 1 | 0 |
| $N(x_1 x_5)$ | 2 | 0 | 0 | $N(x_4 x_4)$ | 4 | 3 | 1 |
| $N(x_1 x_6)$ | 0 | 0 | 0 | $N(x_4 x_5)$ | 1 | 1 | 0 |
| $N(x_1 x_7)$ | 0 | 0 | 0 | $N(x_4 x_6)$ | 0 | 1 | 0 |
| $N(x_2 x_2)$ | 6 | 0 | 0 | $N(x_4 x_7)$ | 0 | 2 | 0 |
| $N(x_2 x_3)$ | 2 | 0 | 0 | $N(x_5 x_5)$ | 3 | 2 | 2 |
| $N(x_2 x_4)$ | 2 | 0 | 0 | $N(x_5 x_6)$ | 0 | 2 | 1 |
| $N(x_2 x_5)$ | 1 | 0 | 0 | $N(x_5 x_7)$ | 0 | 2 | 1 |
| $N(x_2 x_6)$ | 0 | 0 | 0 | $N(x_6 x_6)$ | 0 | 2 | 2 |
| $N(x_2 x_7)$ | 0 | 0 | 0 | $N(x_6 x_7)$ | 0 | 2 | 2 |
| $N(x_3 x_3)$ | 4 | 2 | 1 | $N(x_7 x_7)$ | 0 | 3 | 2 |

The parity matrix *P* is shown below.

| *P* | $x_1$ | $x_2$ | $x_3$ | $x_4$ | $x_5$ | $x_6$ | $x_7$ |
|---|---|---|---|---|---|---|---|
| $x_1$ | 0 | 0 | 0 | 0 | 0 | 0 | 0 |
| $x_2$ | 0 | 0 | 0 | 0 | 1 | 1 | 0 |
| $x_3$ | 0 | 0 | 1 | 1 | 1 | 1 | 1 |
| $x_4$ | 0 | 0 | 1 | 1 | 1 | 1 | 0 |
| $x_5$ | 0 | 1 | 1 | 1 | 1 | 1 | 1 |
| $x_6$ | 0 | 0 | 1 | 1 | 1 | 0 | 0 |
| $x_7$ | 0 | 0 | 1 | 0 | 1 | 0 | 1 |

As $p_{33} = p_{44} = p_{55} = p_{77} = 1$, we construct four test patterns from case (a). The test patterns are: $t_1$(1101111), $t_2$(1110111), $t_3$(1111011) and $t_4$(1111110). Then consider that $p_{66} = 0$ and $p_{65} = 1$, add $t_5$(1111001) (case (b)). No more test case can be derived using case (a) or case (b). Now (case(c)), set $x_1=0$, and obtain $f'_k$s as follows:

$f'_1 = x_2 \oplus x_4 \oplus x_5$     (9)

$f'_2 = x_3 x_4 \oplus x_4 x_7 \oplus x_5 x_6 x_7 \oplus x_3 x_4 x_5 x_6 x_7$     (10)

$f'_3 = x_6 x_7 \oplus x_5 x_6 x_7 \oplus x_3 x_4 x_5$     (11)

The $N_k(x_i x_j)$ values and corresponding $P_1$ matrix are shown below.

| $N_k(x_i x_j)$ | $f_1$ | $f_2$ | $f_3$ | $N_k(x_i x_j)$ | $f_1$ | $f_2$ | $f_3$ |
|---|---|---|---|---|---|---|---|
| $N(x_2 x_2)$ | 1 | 0 | 0 | $N(x_4 x_4)$ | 1 | 3 | 1 |
| $N(x_2 x_3)$ | 0 | 0 | 0 | $N(x_4 x_5)$ | 0 | 1 | 1 |
| $N(x_2 x_4)$ | 0 | 0 | 0 | $N(x_4 x_6)$ | 0 | 1 | 0 |
| $N(x_2 x_5)$ | 0 | 0 | 0 | $N(x_4 x_7)$ | 0 | 2 | 0 |
| $N(x_2 x_6)$ | 0 | 0 | 0 | $N(x_5 x_5)$ | 1 | 2 | 2 |
| $N(x_2 x_7)$ | 0 | 0 | 0 | $N(x_5 x_6)$ | 0 | 2 | 1 |
| $N(x_3 x_3)$ | 0 | 2 | 1 | $N(x_5 x_7)$ | 0 | 2 | 1 |
| $N(x_3 x_4)$ | 0 | 2 | 1 | $N(x_6 x_6)$ | 0 | 2 | 2 |
| $N(x_3 x_5)$ | 0 | 1 | 1 | $N(x_6 x_7)$ | 0 | 2 | 2 |
| $N(x_3 x_6)$ | 0 | 1 | 0 | $N(x_7 x_7)$ | 0 | 3 | 2 |
| $N(x_3 x_7)$ | 0 | 1 | 0 | | | | |

| $P_1$ | $x_2$ | $x_3$ | $x_4$ | $x_5$ | $x_6$ | $x_7$ |
|---|---|---|---|---|---|---|
| $x_2$ | 1 | 0 | 0 | 0 | 0 | 0 |
| $x_3$ | 0 | 1 | 1 | 1 | 1 | 1 |
| $x_4$ | 0 | 1 | 1 | 1 | 1 | 0 |
| $x_5$ | 0 | 1 | 1 | 1 | 1 | 1 |
| $x_6$ | 0 | 1 | 1 | 1 | 0 | 0 |
| $x_7$ | 0 | 1 | 0 | 1 | 0 | 1 |

In $P_1$, as the first entry $p_{11}=1$, from case (a) we derive test pattern $t_6$(0011111). The test set $T_3$ consists of 6 test cases as shown below. The binary tree so formed is shown in Fig.7.

| $T_3$ | $x_1$ | $x_2$ | $x_3$ | $x_4$ | $x_5$ | $x_6$ | $x_7$ |
|---|---|---|---|---|---|---|---|
| $t_1=$ | 1 | 1 | 0 | 1 | 1 | 1 | 1 |
| $t_2=$ | 1 | 1 | 1 | 0 | 1 | 1 | 1 |
| $t_3=$ | 1 | 1 | 1 | 1 | 0 | 1 | 1 |
| $t_4=$ | 1 | 1 | 1 | 1 | 1 | 1 | 0 |
| $t_5=$ | 1 | 1 | 1 | 1 | 0 | 0 | 1 |
| $t_6=$ | 0 | 0 | 1 | 1 | 1 | 1 | 1 |

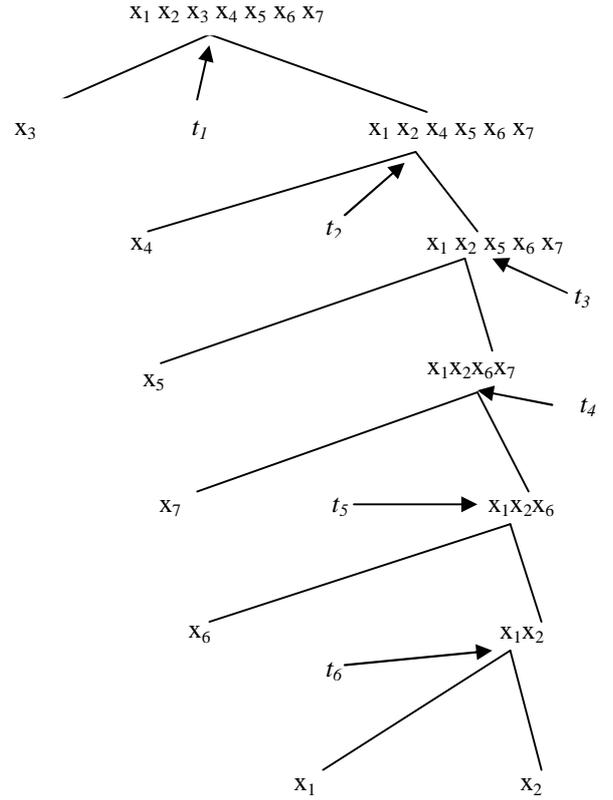

**Fig. 7**: Binary Tree for detection of $(x_i x_j)_+$

### D. Detecting Intra-level Bridging Fault in the EXOR part

Consider a reversible k-CNOT circuit with $p$ observable outputs as shown in Fig.3. Suppose the level of the circuit is $d$. The number of single intra-level bridging faults at a particular level is $n(n-1)/2$ and total number of intra-level SBFs is $(d+1)*n(n-1)/2$. Now consider two cases:

Case 1: $p = 2^k$ for some integer k ≥1. Consider the first test pattern $t_1 = (111…1000…0)$ with $n/2$ consecutive 1s followed by $n/2$ consecutive 0s. The next test pattern is obtained by dividing each block of previous test pattern into two equal length strings of 0s and 1s till we get the last pattern $t_k$ as an alternating sequence of 0s and 1s (see Fig. 8). Clearly $k = log_2 p$ and these k vectors detect all intra-level SBFs at level 0.

Case 2: Let us assume that $2^{(k-1)} < p < 2^k$. We concatenate $(2^k - p)$ bits to $p$ bits and we apply the procedure as in Case 1. At the end, we delete the additional bits from each test pattern. Let's assume the test set so constructed is $T_4$. Hence, size of $T_4$ is $\lceil log_2 p \rceil$.

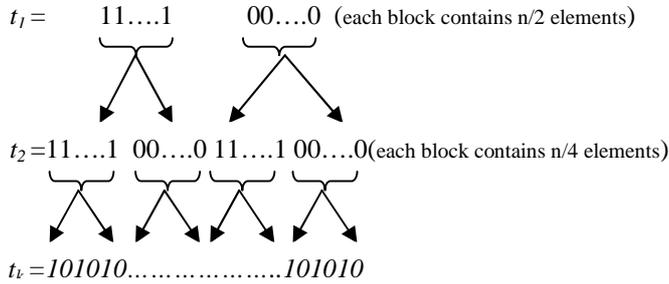

$t_1 = 11\ldots1\ 00\ldots0$ (each block contains n/2 elements)

$t_2 = 11\ldots1\ 00\ldots0\ 11\ldots1\ 00\ldots0$ (each block contains n/4 elements)

$t_k = 101010\ldots\ldots\ldots\ldots101010$

**Fig. 8**: Test generation for SBFs in EXOR part

Consider the k-CNOT circuit as shown in Fig.3. As $p = 3$, we apply case (2) and get two patterns $t_1$ (110) and $t_2$ (101). Now, these 2 patterns detect all SBFs at level 0. If we set $X = (x_1 x_2 \ldots x_n) = (00\ldots0)$, $t_1$ and $t_2$ are transmitted unchanged across all levels. Hence, $t_1$ and $t_2$ detects all intra-level SBFs in the EXOR part.

| $T_4$ | $c_1$ | $c_2$ | $c_3$ | $x_1$ | $x_2$ | $x_3$ | $x_4$ | $x_5$ | $x_6$ | $x_7$ |
|---|---|---|---|---|---|---|---|---|---|---|
| $t_1$ | 1 | 1 | 0 | 0 | 0 | 0 | 0 | 0 | 0 | 0 |
| $t_2$ | 1 | 0 | 1 | 0 | 0 | 0 | 0 | 0 | 0 | 0 |

### E. Detecting $(a_i a_j)_*$ and $(a_i a_j)_+$

Consider the test set $T_5$ shown below. The size of $T_5$ is $n$.

| $T_5$ | $c_1$ | $c_2$ | $c_3$ | $x_1$ | $x_2$ | $x_3$ | $x_4$ | $x_5$ | $x_6$ | $x_7$ |
|---|---|---|---|---|---|---|---|---|---|---|
| $t_1$ | d | d | d | 0 | 1 | 1 | 1 | 1 | 1 | 1 |
| $t_2$ | d | d | d | 1 | 0 | 1 | 1 | 1 | 1 | 1 |
| $t_3$ | d | d | d | 1 | 1 | 0 | 1 | 1 | 1 | 1 |
| $t_4$ | d | d | d | 1 | 1 | 1 | 0 | 1 | 1 | 1 |
| $t_5$ | d | d | d | 1 | 1 | 1 | 1 | 0 | 1 | 1 |
| $t_6$ | d | d | d | 1 | 1 | 1 | 1 | 1 | 0 | 1 |
| $t_7$ | d | d | d | 1 | 1 | 1 | 1 | 1 | 1 | 0 |

where $d \in \{0, 1\}$. Clearly $T_5$ detects all $(a_i a_j)_*$ and $(a_i a_j)_+$. Hence the theorem.

**Theorem 1**: *All single bridging faults (SBFs)(as assumed) in a reversible $(n+p)$-input AND-EXOR based k-CNOT circuit with $p$ observable outputs, can be detected by a test set $T$ where, $T = T_1 \cup T_2 \cup T_3 \cup T_4 \cup T_5$.*

Now, consider the construction of $T_2$. Please note that, every time a new test pattern is generated, the parent subset gets divided into non-empty subsets of smaller size (one of the subsets possibly a single element subset), such that the bridging between inputs from different subsets are detected. Every time a new test pattern is generated, the binary tree is expanded as in Fig. 5. The process of test generation is carried out until all the pendant vertices in Fig. 5 are input variables. Since every test pattern generates a new branch or branches, in the worst case every internal node of the tree corresponds to a test pattern. The number of internal nodes is $(n-1)$ for a binary tree with n pendant vertices and we have for the number of test patterns required to detect all $(x_i x_j)_*$s. Hence, $|T_2| \leq (n-1)$. Similarly it can be shown that $|T_3| \leq (n-1)$. The upper limit of number of test patterns required to detect all SBFs in a reversible k-CNOT circuit can be calculated by summing up the sizes of $T_i$s.

$|T| \leq |T_1| + |T_2| + |T_3| + |T_4| + |T_5|$
$\leq 4 + (n-1) + (n-1) + \lceil log_2 p \rceil + n$
$\leq 3n + \lceil log_2 p \rceil + 2$

**Theorem 2**: *All single bridging faults (SBFs)(as assumed) in a reversible $(n+p)$-input AND-EXOR based k-CNOT circuit with $p$ observable outputs, can be detected by $3n + \lceil log_2 p \rceil + 2$ tests.*

## V. CONCLUSION

In this paper, we have described a simple method for generating a minimal test set for detecting all single bridging faults in an AND-EXOR based reversible k-CNOT circuit. It is observed that the property of reversibility simplifies the test generation problem significantly. Detection of multiple bridging faults (both feedback and non-feedback) in AND-EXOR based reversible k-CNOT circuits would require further investigation.

## REFERENCES

[1] B. Desoete, A. De Vos, M. Sibinski, T. Widerski, "Feynman's reversible logic gates implemented in silicon", Proceedings of the 6th International Conference MIXDES, pp. 496-502, 1999.

[2] P. Picton, "A Universal Architecture for multiple-valued reversible logic", MVL Journal, pp. 27-37, 2000.

[3] K. N. Patel, J. P. Hayes, I. L. Markov, "Fault Testing for Reversible Logic Circuits", Proc. of VTS, pp. 410-416, 2003.

[4] A. Chakraborty, "Synthesis of reversible circuits for testing with Universal Test Set and C-Testability of Reversible iterative logic arrays", proc. VLSI Design, pp. 249-254, 2005.

[5] I. Polian, T. Fiehn, B. Becker, J. P. Hayes, "A Family of Logical Fault Models for Reversible Circuits", proc. 14th Asian Test Symposium, pp. 422-427, 2005.

[6] J. Zhong, J. C. Muzio, "Analyzing Fault Models for Reversible Logic Circuits", IEEE Congress on Evol. Computation, pp. 2422-2427, 2006.

[7] H. Rahaman, D. K. Kole, D. K. Das, B. B. Bhattacharya, "Detection of Bridging Faults in Reversible Circuits", Proc. VLSI Design and Test Symposium, pp. 384-392, 2006.

[8] P. Sarkar, S. Chakrabarti, "Universal test Set for Bridging Fault Detection in Reversible circuit", Proc. Design and Test Workshop, pp. 51-56, 2008.